\documentclass{elsart}
\usepackage{amsfonts}
\usepackage{amssymb}
\usepackage{amsmath}
\usepackage{graphicx,color,psfrag}
\usepackage{url}

\definecolor{dgreen}{rgb}{0,.6,0}

\usepackage[normalem]{ulem}

\begin{document}

\begin{frontmatter}

\title{Breaking an encryption scheme based on chaotic Baker map}

\author[Spain]{Gonzalo Alvarez\corauthref{corr}} and
\author[China]{Shujun Li}
\address[Spain]{Instituto de F\'{\i}sica Aplicada, Consejo Superior de
Investigaciones Cient\'{\i}ficas, Serrano 144---28006 Madrid,
Spain}
\address[China]{Department of Electronic and Information Engineering,
The Hong Kong Polytechnic University, Hung Hom, Kowloon, Hong Kong
SAR, China} \corauth[corr]{The corresponding author, email
address: gonzalo@iec.csic.es.}

\begin{abstract}
In recent years, a growing number of cryptosystems based on chaos
have been proposed, many of them fundamentally flawed by a lack of
robustness and security. This paper describes the security
weaknesses of a recently proposed cryptographic algorithm with
chaos at the physical level. It is shown that the security is
trivially compromised for practical implementations of the
cryptosystem with finite computing precision and for the use of
the iteration number $n$ as the secret key. Some possible
countermeasures to enhance the security of the chaos-based
cryptographic algorithm are also discussed.


\end{abstract}

\end{frontmatter}

\section{Introduction}

In a world where digital communications are becoming ever more
prevalent, there are still services working in analog form. Some
examples of analog communications systems widely used today
include voice communications over telephone lines, TV and radio
broadcasting and radio communications (see
Table~\ref{tab:analog}). Although most of these services are also
being gradually replaced by their digital counterparts, they will
remain with us for a long time. Usually the need to protect the
confidentiality of the information transmitted by these means
might arise. Thus, there is a growing demand for technologies and
methods to encrypt the information so that it is only available in
inteligible form to the authorized users.

In a recent paper \cite{machado04}, a secure communication system
based on the chaotic Baker map was presented, which is a scheme
that encrypts wave signals. First, the analog signal limited in
the bandwidth $W$ is sampled at a frequency $f\geq 2W$ to avoid
aliasing. At the end of the sampling process, the signal is
converted to a sequence
$s^0=\left\{s_1^0,s_2^0,\dots,s_l^0\right\}$ of real values. Next,
the signal is quantized: the amplitude of the signal is divided
into $N$ subintervals and every interval is assigned a real
amplitude value $q_k$, $k=1,\dots,N$, its middle point for
example. Thus, a new sequence is generated by replacing each
$s_i^0$ by the $q_k$ associated to the subinterval it belongs to:
$y^0=\left\{y_1^0,y_2^0,\dots,y_l^0\right\}$, where each $y_i^0$
takes its value from the set $\left\{q_1,\dots,q_N\right\}$. Once
the original wave signal is sampled and quantized, and restricted
to the unit interval, a chaotic encryption signal
$\left\{x_i^0\right\}_{i=1}^l$, $0<x_i^0<1$, is used to generate
the ciphertext. This signal is obtained by either sampling a
chaotic one or by a chaotic mapping. For the purposes of our
analysis, the process to generate the chaotic signal is irrelevant
since our results apply equally to any signal. Finally, an ordered
pair $\left(x_i^0,y_i^0\right)$ is constructed, localizing a point
in the unit square. In order to encrypt $y_i^0$, the Baker map is
applied $n$ times to the point $\left(x_i^0,y_i^0\right)$ to
obtain:
\begin{equation}\label{BakerMap}
\left(x_i^n,y_i^n\right)=\left(2x_i^{n-1} \bmod
1,0.5\left(y_i^{n-1}+\left\lfloor
2x_i^{n-1}\right\rfloor\right)\right).
\end{equation}
The encrypted signal is given by $y_i^n$, where $n$ is considered as
the secret key of the cryptosystem. As a result, a plaintext signal
with values $y_i^0 \in \{q_1,\dots,q_N\}$, is encrypted into a
signal which can take $2^nN$ different values. For a more complete
explanation of this cryptosystem, it is highly recommended the
thorough reading of \cite{machado04}.

In the following two sections, the security defects caused by the
Baker map realized in finite precision are discussed, and then the
fact that the secret key $n$ can be directly deduced from the
ciphertext is pointed out. After the cryptanalysis results, which
constitute the main focus of our paper, some countermeasures are
discussed on how to improve the security of the chaotic
cryptosystem. The last section concludes the paper.

\section{Convergence to zero of the digital Baker map}

\begin{figure}[h]
\center
\includegraphics[width=\textwidth]{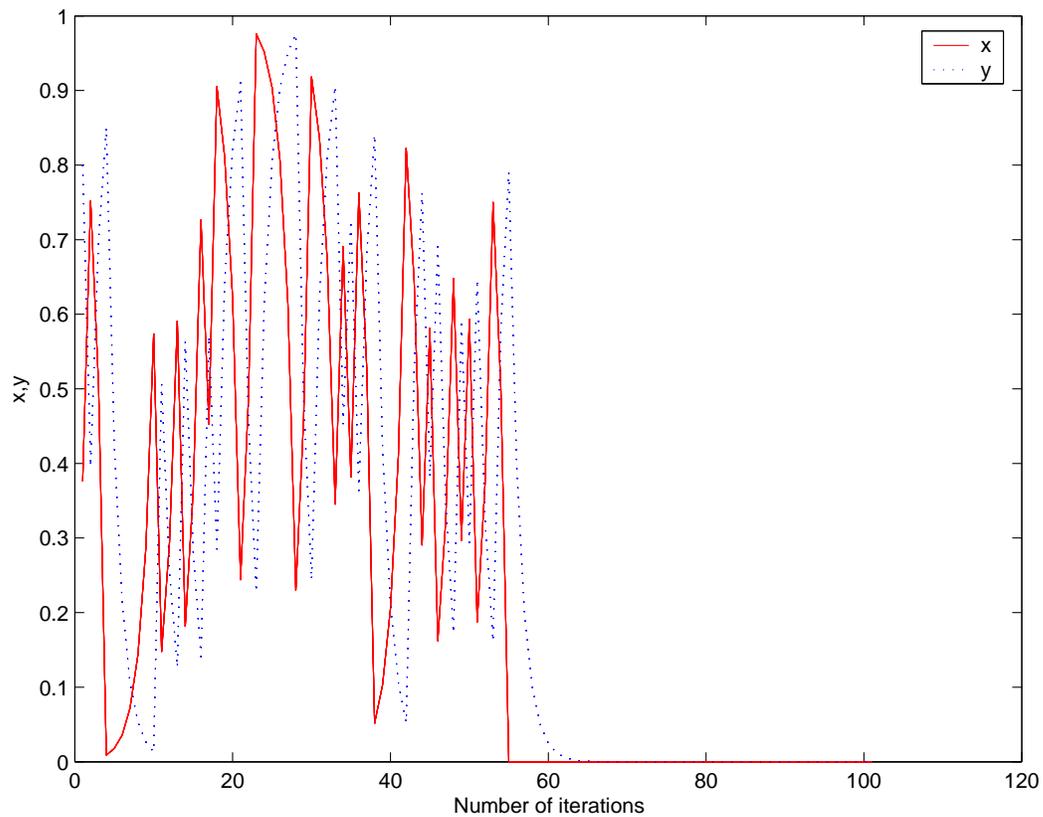}
\caption{\label{fig:map}Orbits followed by $x$ and $y$ in a
practical implementation of the Baker map. As can be observed,
$(0,0)$ constitutes a fixed point. The number of iterations
required to converge to the origin depends on the precision used,
but is always finite in a computer.}
\end{figure}

The proposed cryptosystem uses the Baker map as a mixing function.
The Baker map is an idealized one in the sense that it can only be
implemented with finite precision in digital computers and, as a
consequence, in this case it has a stable attractor at $(0,0)$.
This is easy to see when the value of $x$ is represented in binary
form with $L$ significant bits. Assuming $x^0=0.b_1b_2\cdots
b_j\cdots b_{L-1}b_L$ ($b_j\in\{0,1\}$), the Baker map runs as
follows:
\begin{equation}
x^1=2x^0\bmod 1=x^0\ll1=0.b_2b_3\cdots b_j\cdots
b_{L-1}b_L0,\label{equation:BakerMap}
\end{equation}
where $\ll$ denotes the left bit-shifting operation. Apparently,
the most significant bit $b_1$ is dropped during the current
iteration. As a result, after $m\geq L$ iterations, $x^m\equiv0$.
Once $x^m\equiv0$, it is obvious that $y^j$ will exponentially
converge to zero within a finite number of iterations, i.e., the
digital Baker map will eventually converge to the stable
attractive point at $(0,0)$, as shown in Fig.~\ref{fig:map}. It is
important to note that this result does not depend on the real
number representation method, on the precision, or on the
rounding-off algorithm used, since the quantization errors induced
in Eq. (\ref{equation:BakerMap}) are always zeros in any case.

Considering that in today's digital computers real values are
generally stored following the IEEE floating-point standard
\cite{IEEEStandard754:Floating-Point}, let us see what will happen
when the chaotic iterations run with 64-bit double-precision
floating-point numbers. Following the IEEE floating-point
standard, most 64-bit double-precision numbers are stored in a
normalized form as follows:
\begin{equation}
(-1)^{b_{63}}\times(1.b_{51}\cdots b_0)_2\times 2^{(b_{62}\cdots
b_{52})_2-1023}, \label{equation:Normalized}
\end{equation}
where $b_i$ represent the number bits, $(\cdot)_2$ means a binary
number and the first mantissa bit occurring before the radix dot
is always assumed to be 1 (except for a special value, zero) and
not explicitly stored in $b_{63}\cdots b_0$. When $x^0\in(0,1)$,
assume it is represented in the following format:
\begin{equation}
(1.b_{51}\cdots b_{i+1}1\overbrace{0\cdots 0}^i)_2\times
2^{-e}=(0.\overbrace{0\cdots 0}^{e-1}\overbrace{1b_{51}\cdots
b_{i+1}1}^{53-i})_2,
\end{equation}
where $1\leq e\leq 1022$. Apparently, it is easily to deduce
$L=(e-1)+(53-i)=e+(52-i)$. Considering $0\leq i\leq 52$, $L\leq
1022+52=1074$. When $x^0$ is generated uniformly with the standard
C \texttt{rand()} function in the space of all valid
double-precision floating-point numbers, both $e$ and $i$ will
approximately satisfy an exponentially decreasing distribution,
and then it can be easily proved that the mathematical expectation
of $L$ is about 53 \cite{LiShujun:DigitalChaos2004}.

\begin{figure}[h]
\center
\psfrag{Error}{Error}\psfrag{n}{$n$}\includegraphics[width=\textwidth]{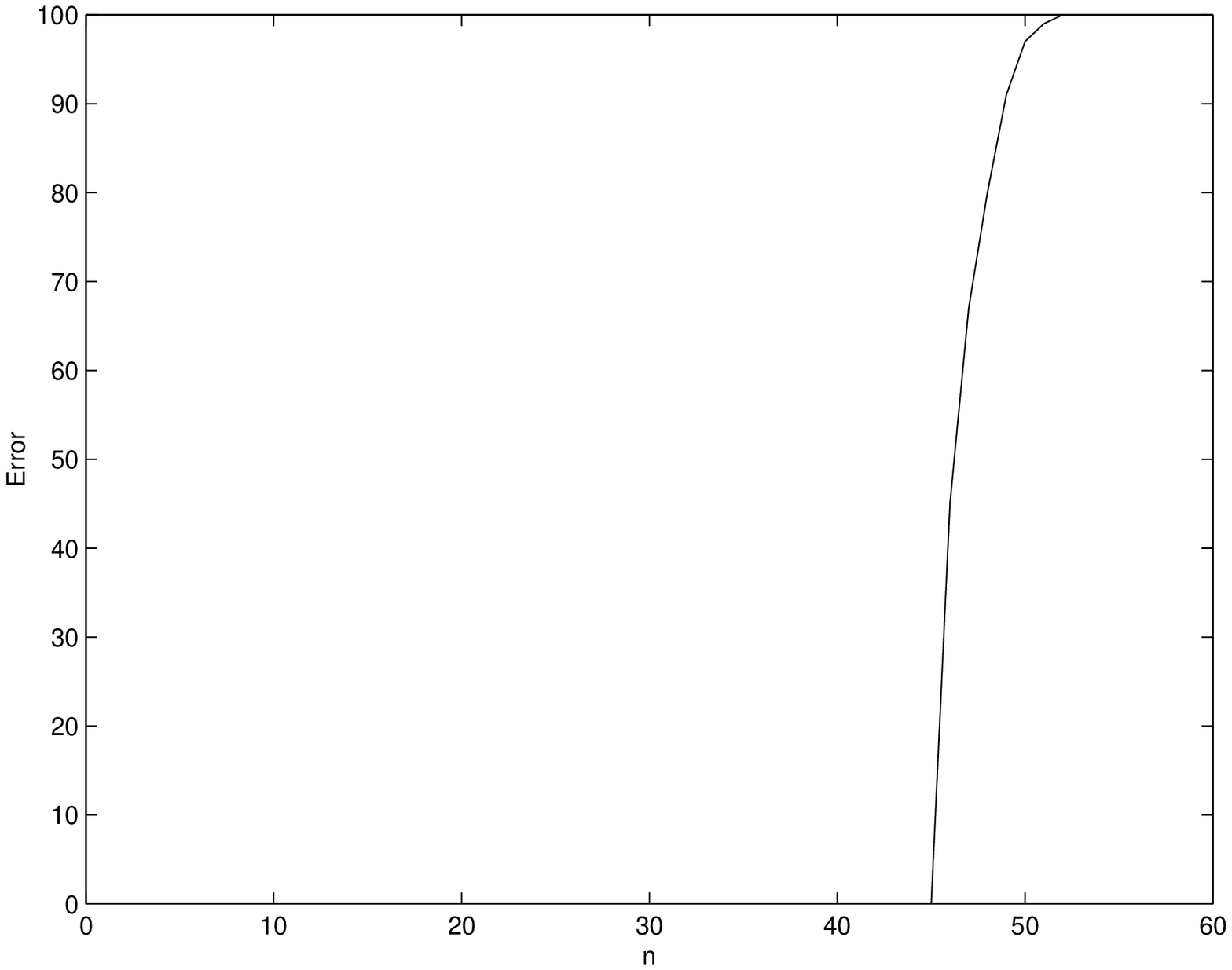}
\caption{\label{fig:ber}Number of errors when decrypting a
100-sample signal for different values of the secret key $n$ using
double-precision floating-point arithmetic.}
\end{figure}

This means that the value of the secret key $n$ must not be
greater than 53. In other words, it is expected that each
plaintext sample $y_i^0$ cannot be correctly decrypted when $n$ is
greater than 53 (or even smaller but close to 53), since the
counter-iterating process is unable to get $x_i^0$ from $x_i^n=0$
due to the loss of precision during the forward iterations.
Figure~\ref{fig:ber} plots the recovery error obtained for
different values of the secret key $n$ when a 100-sample
ciphertext is decrypted. It can be appreciated how the plaintext
is correctly recovered only when $n\leq 45$. For $n\geq 52$, the
system does not work at all. As a consequence, only $n=45$ secret
keys have to be tried to break a ciphertext encrypted with this
cryptosystem. This takes a modern desktop computer less than a
second for moderated lengths of the plaintext. This attack is
called a brute force attack, which breaks a cipher by trying every
possible key. The feasibility of a brute force attack depends on
the size of the cipher's key space and on the amount of
computational power available to the attacker. With today's
computer technology, it is generally agreed in the cryptography
community that a size of the key space $K<2^{100}\approx 10^{30}$
is insecure \cite{ctap}. Compare this figure with the key space
$K=45$ of the cipher under study.

If the value of $n$ could be arbitrarily enlarged, then the
encryption process would slow down until it would be unusable in
practice. Thus, from any point of view, this is an impractical
encryption method because it is either totally insecure or
infinitely slow, without any reasonable tradeoff possible. In
\cite{machado04} it is said that the encryption is applied to the
wave signal instead of the symbolic sequence. Therefore, in
Table~\ref{tab:analog} a review of some widely used multimedia
communications systems with their bandwidth and sampling
frequencies is given. These are the kind of signals that might be
encrypted by the system proposed in \cite{machado04}. Consider for
example TV broadcasting, which transmits 12,000,000 samples per
second. It is impossible to iterate the Baker map billions of
times for 12,000,000 samples in one second with average computing
power.

Finally, another physical limitation of the cryptosystem is that
when $n$ is very large, each encrypted sample $y_i^n$ would
require a vast amount of bits to be transmitted, which would
require in turn a transmission channel with infinite capacity,
meaning that the system cannot work in practice.

\section{Determinism of the ciphertext}

Even assuming that the messages are encrypted with an imaginary
computer with infinite precision and infinite speed, using an
infinite-bandwidth channel, and an idealized version of the Baker
map, the cryptosystem would be broken as well because the secret
key $n$ can still be derived from only one amplitude value of the
ciphertext. To begin with, let us assume that two quantization
levels are used, that is, $N=2$. During the encryption process a
binary tree is generated in the following way:
\begin{equation}\label{Eq:tree}
  y_i^0=\left\{\begin{array}{l}
 0.25\;(0.01)_2\rightarrow y_i^1=\left\{ \begin{array}{l}
 0.125\;(0.001)_2 \rightarrow y_i^2=\left\{ \begin{array}{l}
 0.0625\;(0.0001)_2\\
 0.5625\;(0.1001)_2 \\
 \end{array} \right. \\
 0.625\;(0.101)_2\rightarrow y_i^2=\left\{ \begin{array}{l}
 0.3125\;(0.0101)_2\\
 0.8125\;(0.1101)_2 \\
 \end{array} \right. \\
 \end{array} \right. \\
 0.75\;(0.11)_2\rightarrow y_i^1=\left\{ \begin{array}{l}
 0.375\;(0.011)_2\rightarrow y_i^2=\left\{ \begin{array}{l}
 0.1875\;(0.0011)_2 \\
 0.6875\;(0.1011)_2 \\
 \end{array} \right. \\
 0.875\;(0.111)_2\rightarrow y_i^2=\left\{ \begin{array}{l}
 0.4375\;(0.0111)_2 \\
 0.9375\;(0.1111)_2 \\
 \end{array} \right. \\
 \end{array} \right. \\
 \end{array} \right.,
\end{equation}
where $(\cdot)_2$ following the decimal number denotes its binary
format. The fact that the ciphertext uses $2^nN$ discrete
amplitudes constitutes its weakest point. It is possible to
directly get the value of $n$ with only one known amplitude. In
Eq.~(\ref{Eq:tree}), it is obvious that $y_i^n$ is always one
value in the set
\begin{equation}\label{eq:set}
\left\{\dfrac{2j+1}{2^{n+2}}\right\}_{j=0}^{j=2^{n+1}-1}=\left\{\dfrac{1}{2^{n+2}},\cdots,
\dfrac{2^{n+2}-1}{2^{n+2}}\right\}.
\end{equation}
As mentioned above, in the case that the real values are stored in
the IEEE-standard floating-point format
\cite{IEEEStandard754:Floating-Point}, any amplitude value $y_i^n$
will be represented in the following form:
\begin{equation}
y_i^n=+1.b_1b_2\cdots b_l\times 2^{-e}=0.\overbrace{0\cdots
0}^{e-1}1b_1b_2\cdots b_l,
\end{equation}
where $b_l=1$. From Eq.~(\ref{eq:set}), one can see that
$l+e=n+2$. Therefore, we can directly derive $n=(l+e)-2$, by
checking which bit is the least significant bit (i.e., the least
significant 1-bit) in all bits of $y_i^n$.

A more intuitive way to compute $n$ from a single amplitude value,
$y_i^n$, consists of two steps: i) represent this amplitude value
in fixed-point binary form; ii) count the bits in the fixed-point
format of $y_i^n$ to determine the value of an integer $B$, which
is the number of bits after the radix dot and before the least
significant bit, i.e., $y_i^n=0.\underbrace{\overbrace{0\cdots
0}^{e-1}1b_1b_2\cdots b_l}_{B=l+e}0\cdots 0$. Obviously, $n =
B-2$. Similarly, for other values of $N=2^v$, one can easily
deduce that $n=(l+e)-(v+1)=B-(v+1)$; and for $N\neq 2^v$, the
value of $n$ can still be derived easily, but the calculation
algorithm depends on how the binary tree shown in
Eq.~(\ref{Eq:tree}) is re-designed.

Although in \cite{machado04} it is hinted that the value of $n$
could be changed dynamically based on some information of the
encrypted trajectory, this idea would not further increase the
security of the cryptosystem as long as $2^nN$ different
amplitudes are still possible for each different $n$ value. This
means that the ciphertext value $y_i^{n_i}$, whatever $n_i$, can
only take values from the finite set defined in Eq.~(\ref{eq:set})
for the given $n_i$. Hence, for each $y_i^{n_i}$ the value of
$n_i$ can be computed as described above and the security is again
compromised.

\section{Some possible countermeasures of enhancing the cryptosystem}

There are many ways to improve the security of the attacked
cryptosystem. This section introduces three possible ones:
changing the key, changing the 2-D chaotic map, and masking the
ciphertext with a secret signal. Note that only the basic ideas
are given, and the concrete designs and detailed security analysis
are omitted because this is not the main focus of our paper.

\subsection{Changing the key}

As mentioned above, in addition to the above-discussed security
defects of the secret key $n$, using $n$ as the secret key has
another obvious paradox: from the point of view of the security,
$n$ should be as large as possible; while from the point of view
of the encryption speed, $n$ should be as small as possible.
Apparently, $n$ is not a good option as the secret key.

Instead of using $n$, better candidates for the secret key must be
chosen, such as the control parameter of the 2-D chaotic map and
the generation parameter of the encryption signal $x$. If the
former is chosen, the Baker map has to be modified to introduce
some secret control parameters, as described in the following
section.

\subsection{Changing the 2-D chaotic map}

As shown above, the multiplication factor 2 in the original Baker
map is the essential reason of its convergence to $(0,0)$ in the
digital domain, so the Baker map has to be modified to cancel this
problem, or another 2-D chaotic map without this problem has to be
used.

A possible way is to generalize the original Baker map to a
discretized version over a $M\times N$ lattice of the unit plane.
For example, when $M=N=2$, the lattice is composed of the
following four points: $(0.125,0.125)$, $(0.125,0.725)$,
$(0.725,0.125)$ and $(0.725,0.725)$. A typical example of Baker
map discretized in this way can be found in
\cite{Fridrich:IJBC98}, reproduced next for convenience.

First, the standard Baker map is generalized by dividing the unit
square into $k$ vertical rectangles, $[F_{i-1},F_i)\times [0,1)$,
$i=1,\dots,k$, $F_i=p_1+p_2+\dots+p_i$, $F_0=0$, such that
$p_1+\dots+p_k=1$. The lower right corner of the $i$th rectangle
is located at $F_i=p_1+\dots+p_i$. Formally the generalized map is
defined by:
\begin{equation}\label{GeneralizedBaker}
    B_c(x,y)=\left(\frac{1}{p_i}(x-F_i),p_iy+F_i\right),
\end{equation}
for $(x,y)\in[F_i,F_i+p_i)\times[0,1)$.

The next step consists of discretizing the generalized map. If one
divides an $N\times N$ square into vertical rectangles with $N$
pixels high and $N_i$ pixels wide, then the discretized Baker map
can be expressed as follows:
\begin{equation}\label{DiscretizedBaker}
B_d(r,s)=\left(\frac{N}{n_i}(r-N_i)+\left(s\bmod\frac{N}{n_i}\right),\frac{n_i}{N}\left(s-\left(s\bmod\frac{N}{n_i}\right)\right)+N_i\right),
\end{equation}
where the pixel $(r,s)$ is with $N_i\leq r<N_i+n_i$, $0\leq s<N$.
The sequence of $k$ integers, $n_1,n_2,\dots,n_k$, is chosen such
that each integer $n_i$ divides $N$, and $N_i=n_1+n_2+\dots+n_k$.
The formula can be extended for $M\times N$ rectangles (see
\cite{Fridrich:IJBC98}).

With such a discretization, the negative convergence to zero can
be removed. However, another negative digital effect, the
recurrence of the orbit, arises in this case, since any orbit will
eventually become periodic within $MN$ iterations. This means that
the security defect caused by the small key space is not
essentially improved. Thus, the discretized Baker map must be used
when the key is changed to be its discretization parameters.

Another way is to use entirely different 2-D chaotic maps with one
or more adjustable parameters, which can be used as the secret key
instead of $n$.

\subsection{Masking the ciphertext with a secret pseudo-random signal}

An easy way to enhance the security of the cryptosystem is to mask
the ciphertext with a \textit{secret} pseudo-random signal, which
can efficiently eliminate the possibility to derive the estimated
value of $n$ from one amplitude of the ciphertext. The secret
masking sequence can be the chaotic encryption signal $\{x_i^0\}$,
and the parameters of controlling the generation process of
$\{x_i^0\}$ should be added as part of the secret key. In this
case, the ciphertext is changed from $\{y_i^n\}$ into
$\{y_i^n+x_i^0\}$. Note that the masking can be considered as an
added stream cipher to the original system. This is a common
technique to achieve stronger ciphers \cite{LiShujun:PLA2003}.

\section{Conclusions}
\label{sec:conclusion}

In summary, the new cryptosystem proposed in \cite{machado04} can
be broken due to the limitation of computers to represent real
numbers. Even if an ideal computer with infinite precision were
used to encrypt the messages, the cipher can still be broken due
to the fact that the number and value of possible amplitude values
in the ciphertext depend directly on the secret key $n$.
Furthermore, for the cryptosystem to work with large values of
$n$, an ideal computer with infinite computing speed, infinite
storage capacity, and infinite transmission speed would be
required. As a consequence, we consider that this cryptosystem
should not be used in secure applications. Some possible
countermeasures are also discussed on how to improve the security
of the cryptosystem under study. An important conclusion of our
work is that an idealized map cannot be used in a practical
implementation of a chaos-based cipher.

\ack{This work is supported by Ministerio de Ciencia y
Tecnolog\'{\i}a of Spain, research grant TIC2001-0586 and
SEG2004-02418. We also thank the anonymous reviewers for their
valuable suggestions.}

\clearpage \pagestyle{empty}

\section*{Tables}

\begin{table}[h]
  \centering
  \caption{Multimedia communication systems and their bandwidth.}
  \label{tab:analog}
  \begin{tabular}{lrr}
  \hline
    Communication system & Bandwidth (KHz) & Sampling frequency (Ksamples/s) \\
  \hline
    Voice over telephone & 3.3 & 8 \\
    Radio communications & 3.3 & 8 \\
    Radio Broadcast (AM) & 5 & 10 \\
    Radio Broadcast (FM) & 15 & 30 \\
    TV & 5500 & 12000 \\
  \hline
  \end{tabular}
\end{table}

\clearpage

\section*{Figures}

\begin{thebibliography}{1}

\bibitem{machado04}
R.~F. Machado, M.~S. Baptista, and C.~Grebogi.
\newblock Cryptography with chaos at the physical level.
\newblock {\em Chaos, Solitons and Fractals}, 21(5):1265--1269, 2004.

\bibitem{IEEEStandard754:Floating-Point}
IEEE~Computer Society.
\newblock {IEEE} standard for binary floating-point arithmetic.
\newblock ANSI/IEEE Std. 754-1985, August 1985.

\bibitem{LiShujun:DigitalChaos2004}
S. Li.
\newblock When chaos meets computers.
\newblock arXiv:nlin.CD/0405038, available online at
\url{http://arxiv.org/abs/nlin/0405038}, May 2004.

\bibitem{ctap}
D.~R. Stinson.
\newblock {\em Cryptography: Theory and Practice}.
\newblock CRC Press, 1995.

\bibitem{Fridrich:IJBC98}
J.~Fridrich.
\newblock Symmetric ciphers based on two-dimensional chaotic maps.
\newblock {\em Int. J. Bifurcation and Chaos}, 8(6):1259--1284, 1998.

\bibitem{LiShujun:PLA2003}
S.~Li, X.~Mou, Z.~Ji, J.~Zhang, and Y.~Cai.
\newblock Performance analysis of {Jakimoski-Kocarev} attack on a class of
  chaotic cryptosystems.
\newblock {\em Physics Letters A}, 307(1):22--28, 2003.

\end{thebibliography}
\end{document}